\documentstyle[epsf,12pt]{article}

\begin{document}
%\tableofcontents
%\chapter{INTRODUCTION}
\begin{titlepage}
\centerline{\large{\bf{Warm Asymmetric Nuclear Matter 
                                        and Proto-Neutron Star}}} 
\vspace{0.5in}
\centerline { Pravat Kumar Jena$^*$ and Lambodar Prasad Singh$^+$}
%\date{\today}
\vspace{0.25in}
\centerline{Department of Physics, Utkal University, Vanivihar,}
\centerline{ Bhubaneswar-751004, India.}
%\maketitle
\vspace{1in}
%\begin{abstract}
\centerline{\bf{Abstract}}
\vspace{.1in}

  Asymmetric nuclear matter equation of state at finite temperature is studied
in SU(2) chiral sigma model using mean field approximation. The effect of 
temperature on effective mass, entropy, and binding energy is
discussed. Treating the system as one with two conserved charges the
liquid-gas phase transition  is  investigated. We have also discussed
the effect of proton fraction on critical temperature with and without 
$\rho$-meson contribution. We have extended our work to study the
structure of proto-neutron star with neutron free charge-neutral
matter in beta-equilibrium. We found that the mass and radius of the
star decreases as it cools from the entropy per baryon S = 2 to S = 0 and      
the maximum temperature of the core of the star is about 62 MeV for S = 2.

%\end {abstract}
\vspace{0.5in}

 PACS Nos: 26.60.+c, 21.30.Fe, 21.60.Jz . 

\vspace{1.2in}
\noindent
%\hline
$^*$email: pkjena@iopb.res.in \\
$^+$email: lambodar@iopb.res.in

\end{titlepage}

\section{Introduction}
  The study of properties of hot dense asymmetric nuclear matter has been, in 
the last few years, vigorously pursued  in connection with
astrophysical problems[1,2], such as supernova explosions, the evolution of
neutron stars and nuclear problems such as heavy-ion-collisions. As the 
equation of state(EOS) describes the variation of energy density and pressure 
with density  and temperature, it can be used  to  describe different
phases like gaseous and liquid nuclear matter  upto the 
deconfinement transition. It is also possible to study the liquid-gas
phase transition, which may occur in the warm and dilute nuclear
matter produced in heavy-ion collisions. Several authors using  
non-relativistic[3] and relativistic[4-8] theories have studied   
liquid-gas phase transition. Most of the calculations found the
critical temperature $T_c$ lying in the range of 14-20 MeV for
symmetric  nuclear matter. The critical temperature of symmetric
nuclear matter for the most acceptable relativistic Walecka model, 
is $T_c \approx$ 18.3 MeV[9]. As the asymmetric parameter or the proton  
fraction plays a vital role in getting  the critical temperature, the 
addition of $\rho$-meson is  quite essential for the study of asymmetric  
nuclear matter[4,5].

     Field theoretical finite temperature EOS
plays an important role in studying the properties and structure of
hot dense massive stars such as proto-neutron 
stars at different  temperatures and densities. The important 
characteristics which determine the composition of matter in a compact 
star are[10] their relative  compressibilities(important to determine 
maximum mass of neutron star), symmetry energies( important to determine 
the typical stellar radius and the relative n, p, e, neutrino
abundances) and specific heats(important to determine local temperatures).   
These characteristics vary with respect to the  EOS used in different
models. Generally the structures of both hot and cold, and both  
neutron-rich and neutron-poor, stars are fixed by the EOS[2].
A Proto-Neutron Star(PNS) is born following the gravitational collapse
of the core of a massive star during a supernova explosion(type-II) and
evolves to a cold and deleptonized neutron star, basically through neutrino
emission. This very dense and hot core is also able to trap neutrinos, 
imparting momentum to the outer layers and then cooling as it reaches a 
quasi-equilibrium state. There can also be a quark-hadron phase transition 
in PNS at high density and temperature[11]. 

   In this paper we have extended our earlier study[12] on finite
temperature EOS by adding $\rho$-meson to the Lagrangian density
within the MCH model[13] as the addition of $\rho$-meson is quite essential 
for the study of  asymmetric nuclear matter. In our earlier work we have 
studied the effect of temperature on EOS, effective mass, entropy and binding
energy for symmetric nuclear matter and investigated the liquid-gas
phase transition using the same model without the $\rho$-meson. We now study 
here how the asymmetric parameter or proton
fraction in addition to $\rho$-meson changes those properties at
finite temperatures. As finite temperature EOS has an important role in 
studying the structure and properties of astrophysical objects, we
have also investigated the structure of PNS taking $\beta$-stable charge
neutral matter without neutrinos consisting of neutrons, protons and
electrons only.
   
  In Sec.2, asymmetric nuclear matter at finite temperature is
presented. Proto-neutron star is discussed in Sec.3. We conclude with
some remarks in Sec.4.

\section{\bf{Asymmetric nuclear matter at finite temperature}}
\subsection{Equation of state}

    Continuing our earlier investigation [12]  on the study of finite 
temperature effect of asymmetric nuclear matter using MCH model we 
include here the $\rho$-field in the Lagrangian density. 
The EOS for hadronic phase is calculated by using the  Lagrangian density[13] 
(with $\hbar=c=K_{Boltzmann}=1$),
\begin{eqnarray}
 L = \frac{1}{2}(\partial_{\mu} \vec{\pi}.\partial^{\mu}\vec{\pi} + 
   \partial_{\mu}\sigma  
  \partial^{\mu}\sigma )-\frac{1}{4}F_{\mu \nu} F_{\mu \nu}-
   \frac{\lambda}{4}(x^2-x_0^2)^2 
  -\frac{\lambda B}{6m^2}(x^2-x_0^2)^3 \nonumber \\ 
   -\frac{\lambda C}{8m^4}(x^2-x_0^2)^4 - g_{\sigma}\bar{\psi }(\sigma +
  i\gamma_{5}\vec{\tau} .\vec{\pi} )\psi  
 +\bar{\psi}(i \gamma_{\mu}
  \partial ^{\mu} -g_{\omega}\gamma_{\mu}\omega ^{\mu})\psi \nonumber \\   
   +\frac{1}{2}g_{\omega}^2  x^2 \omega_{\mu}\omega ^{\mu} 
  -\frac{1}{4} G_{\mu \nu}.G^{\mu:q
\nu}+\frac{1}{2}
   m_{\rho}^{2}\vec{\rho_{\mu}}.\vec{\rho^{\mu}}
   -\frac{1}{2}g_{\rho}\bar{\psi}(\vec{\rho_{\mu}}.\vec{\tau}\gamma^{\mu})\psi 
\end{eqnarray}
\noindent
In the above Lagrangian, $F_{\mu \nu} \equiv \partial_{\mu}\omega_{\nu} 
 - \partial_{\nu} \omega_{\mu}$, $G_ {\mu \nu} \equiv \partial_{\mu}\rho_{\nu} 
 - \partial_{\nu} \rho_{\mu}$ and $x = (\vec{\pi}^2 +\sigma ^2)^{1/2}$, 
$\psi $ is the nucleon  isospin doublet, $\vec{\pi}$ is the 
pseudoscalar-isovector pion field, $\sigma$ is the scalar field and 
$\omega_{\mu}$, is a dynamically generated isoscalar vector field, which
couples to the conserved baryonic current
$j_{\mu}=\bar{\psi}\gamma_{\mu}\psi$. $\vec{\rho_{\mu}}$ is the
isotriplet vector meson field with mass $m_{\rho}$. B and C are constant  
coefficients associated  with the higher order self-interactions of the 
scalar field .

   The masses of the nucleon, the  scalar meson  and the vector meson
are respectively given by 
\begin{equation}
  m = g_{\sigma}x_0, \  m_{\sigma} =\sqrt{2\lambda} x_0, \ 
 m_{\omega} = g_{\omega}x_0 
\end{equation}
\noindent
   Here $x_0$ is the vacuum expectation value of the  $\sigma $ field , 
$ g_{\omega}$, $g_{\rho}$ and $g_{\sigma}$
are the coupling constants for the vector and scalar fields respectively  
 and $\lambda =
(m_{\sigma}^2 - m_{\pi}^2)/ (2 \it {f}_{\pi}^2) $, where  $m_{\pi}$ is the 
pion mass , $\it{f}_{\pi}$  is the pion decay coupling constant .

   Using Mean-field  approximation, the equation of motion for isoscalar 
vector meson  field  is 
\begin{equation}
   \omega_{0}=\frac{n_{B}}{g_{\omega}x^2}
\end{equation}
\noindent 
and that for  $\rho$-field is given by  
\begin {equation}
 \rho_{0}^{3} = (g_{\rho}/2m_{\rho}^{2})(n_p - n_n)
\end{equation} 
\noindent 
$n_B(=n_{p}+n_{n})$ is the baryon number density at temperature T 
and is given by
\begin{equation}
n_B = \frac{\gamma}{(2\pi)^3}\int_{0}^{\infty} d^3k \left[ n_{i}(T) 
 - \bar n_{i}(T)\right]
\end{equation}
\noindent 
with 
\begin{eqnarray}
  n_i(T) = \frac{1}{e^{( E^*(k) - \nu_i )\beta} + 1} , \\
\bar n_i(T) = \frac{1}{e^{( E^*(k) + \nu_i )\beta} + 1} ,
\end{eqnarray}
\noindent 
where $i = n,p$. $E^*(k) = (k^2 + y^2m^2)^{1/2}$ is the effective nucleon
energy, $\beta = 1/K_BT$, $\gamma$ is the spin degeneracy factor with
$n_i(T)$ and $\bar n_i(T)$ being  Fermi-Dirac distribution functions
for particle and  anti-particle respectively at finite temperature.  
$\nu_i$ is the effective  baryon chemical potential which is related
to the chemical potential $\mu_i$ as 
\begin{eqnarray}
\nu_p = \mu_{p} - \frac{C_{\omega}n_B}{y^2} - \frac{C_{\rho}}{4}(n_p - n_n)\\
\nu_n = \mu_{n} - \frac{C_{\omega}n_B}{y^2} + \frac{C_{\rho}}{4}(n_p - n_n)
\end{eqnarray}
\noindent 
and $y \equiv x/x_0 $ is the effective mass factor which must be 
determined self consistently from the equation of motion for scalar field 
which is given by
\begin{equation}
 (1-y^2)-\frac {B}{m^2C_{\omega}}(1-y^2)^2 +\frac {C}{m^4C_{\omega}^2}
     (1-y^2)^3 +\frac{2 C_{\sigma}C_{\omega}n_{B}^2}{m^2y^4}
   -\frac{C_{\sigma}\gamma}{\pi^2} 
\int_{0}^{\infty} \frac{dk k^2 (n_i(T) + \bar n_i(T))}{\sqrt{k^2+{m^*}^2}} = 0.
\end{equation}
\noindent
 $m^* \equiv ym$ is the effective mass of the nucleon and the coupling 
constants are 
\begin{equation}
   C_{\sigma} \equiv  \frac {g_{\sigma}^2}{m_{\sigma}^2},\ \ 
   C_{\omega}\equiv \frac {g_{\omega}^2}{m_{\omega}^2} \nonumber \ \ 
and \ \ C_{\rho} \equiv g_{\rho}^2/m_{\rho}^2 \ .
\end{equation} 
\noindent
 The symmetric energy coefficient that follows from the semi-empirical
nuclear mass formula is[13]  
\ \ \ \ \  $ a_{sym} = \frac{C_{\rho}k_{f}^{3}}{12 \pi^{2}}
   +\frac{k_f^{2}}{6 \sqrt{k_f^2 +{m^*}^2}}$ . \\

   Now the nucleon number densities, energy density and pressure at finite
temperature and finite density are given by
\begin {equation}
n_p = \frac{\gamma}{(2\pi)^3}\int_{0}^{\infty} d^3k \left[ n_{p}(T) 
 - \bar n_{p}(T)\right]
\end{equation}
\begin {equation}
n_n = \frac{\gamma}{(2\pi)^3}\int_{0}^{\infty} d^3k \left[ n_{n}(T) 
 - \bar n_{n}(T)\right]
\end{equation}
\noindent
\begin{eqnarray}
\epsilon = \frac{m^2(1-y^2)^2}{8C_{\sigma}}-\frac{B}{12C_{\omega} C_{\sigma}}
 (1-y^2)^3 +\frac{C}{16m^2C_{\omega}^2C_{\sigma}}(1-y^2)^4 
+\frac{C_{\omega}n_B^2}{2y^2} + \frac{1}{2}m_{\rho}^2(\rho_{0}^3)^2\nonumber\\ 
+ \frac{\gamma}{2\pi^2} \int_{0}^{\infty} dk 
k^2\sqrt{(k^2+{m^*}^2)}\left[n_n(T) + \bar n_n(T) + n_p(T) + \bar n_p(T)\right]
\end{eqnarray}
\noindent
\begin{eqnarray}
P = -\frac{m^2(1-y^2)^2}{8C_{\sigma}} +\frac {B}{12C_{\omega} C_{\sigma}}
   (1-y^2)^3-\frac{C}{16m^2C_{\omega}^2C_{\sigma}}(1-y^2)^4  
        +\frac{C_{\omega}n_B^2}{2y^2} \nonumber \\ 
+ \frac{1}{2}m_{\rho}^2 (\rho_{0}^3)^2+\frac{\gamma}{6\pi^2}\int_{0}^{\infty}
 \frac{dk k^4[n_n(T) + \bar n_n(T) + n_p(T) + \bar n_p(T)]}
   {\sqrt{(k^2+{m^*}^2)}}
\end{eqnarray}
\noindent
The entropy density(S/V) and entropy per baryon(S) can be obtained as 
\begin {equation}
 S/V = (P + \epsilon -\mu_p n_p-\mu_n n_n)\beta 
\end{equation} 
\begin {equation}
 S = (P + \epsilon -\mu_p n_p-\mu_n n_n)\beta/n_B
\end{equation} 
\noindent
In order to describe the asymmetric nuclear matter one can introduce
the proton fraction which is defined as
\begin {equation}
 y_p = \frac {n_p}{n_B}
\end{equation} 
\noindent
For the neutron matter $y_p = 0$ and for symmetric nuclear matter $y_p = 0.5$.

    The values of four parameters $C_\sigma, C_{\omega}, C_{\rho}$, B and
C occurring in the above equations are obtained[13]  by fitting with the 
saturation values of binding energy/nucleon(-16.3 MeV),   
the saturation density(0.153 fm$^{-3}$), the symmetric
energy(32 MeV), the effective(Landau) mass(0.85 M), and 
nuclear incompressibility($\sim $300 MeV),   
are  $C_{\omega}$ = 1.999 fm$^2$, $C_{\sigma}$ = 6.816 
fm$^2$, $C_{\rho}$ = 4.661 fm$^2$, B = -99.985 and C = -132.246. 
For  a given value of $n_B$ at fixed $y_p$ and/or T, the
equations[10,12,13] can be solved self-consistently to get
y, $\mu_p$, $\mu_n$ and making substitution, the energy density,
pressure, entropy density and entropy per baryon can be evaluated.

  Now let us discuss the liquid-gas phase transition. In case of the
more common single component phase equilibria, such as liquid vapour,
the phases are distinguished by only one parameter, e.g. the density
whereas the binary mixture has the additional parameter, the proton
fraction $y_p$, which is different from one phase to the other[14].
For the description of liquid-gas phase transition we have followed here the 
thermodynamic approach of Refs.[7,14,15]. In the case of asymmetric nuclear 
matter the system is characterized by two conserved charges namely the baryon 
density($n_B = n_p +n_n$) and the total charge or equivalently the third 
component of isospin($I_3 = \frac{n_p-n_n}{2} $) or equivalently the proton 
fraction($y_p$). Thus the stability criteria may be expressed by the following
relations[7]
\begin{equation}
n_B\left(\frac{\partial{P}}{\partial{n_B}}\right)_{T,y_p} > 0
\end{equation}

\begin{eqnarray}
\left(\frac{\partial{\mu_p}}{\partial{y_p}}\right)_{T,P} > 0, \ \ or \ \ 
\left(\frac{\partial{\mu_n}}{\partial{y_p}}\right)_{T,P} < 0 
\end{eqnarray}
\noindent
The first inequality shows that isothermal compressibility is positive which
implies that the system is mechanically stable. The second condition expresses
the diffusive stability. If any of the three stability criteria is violated 
then there will be a phase separation[7].

\subsection{\bf{Results and discussions }}

    First, we discuss the effective nucleon mass of nuclear
matter at finite temperature and density. Fig.1 shows the effective mass 
versus number density at temperatures T=0, 100, 200 MeV. The solid lines 
are for $y_p=0.5$ and dotted lines  are for $y_p=0$. It is observed
that in all temperatures $M^*$ decreases with the increase of
$n_B$. For T=0 or T=100 MeV, the effective mass varies little with
$y_p$, whereas it is very sensitive with $y_p$ for T=200 MeV.
It is also clear that at zero density, the effective mass is the same
for neutron matter and symmetric matter at different
temperature. Hence these result indicate that the nucleons of neutron
matter must stay in higher energy levels compared to that of symmetric matter
in order to have the same number density and the difference will
decrease as the number density decreases. This result  agrees with
that obtained in Ref.6. 
\begin{figure}[t]
\leavevmode
%\centerline{\psfxsize=2.5in\psfbox{pmu3.ps}}
\protect\centerline{\epsfxsize=5in\epsfysize=5in\epsfbox{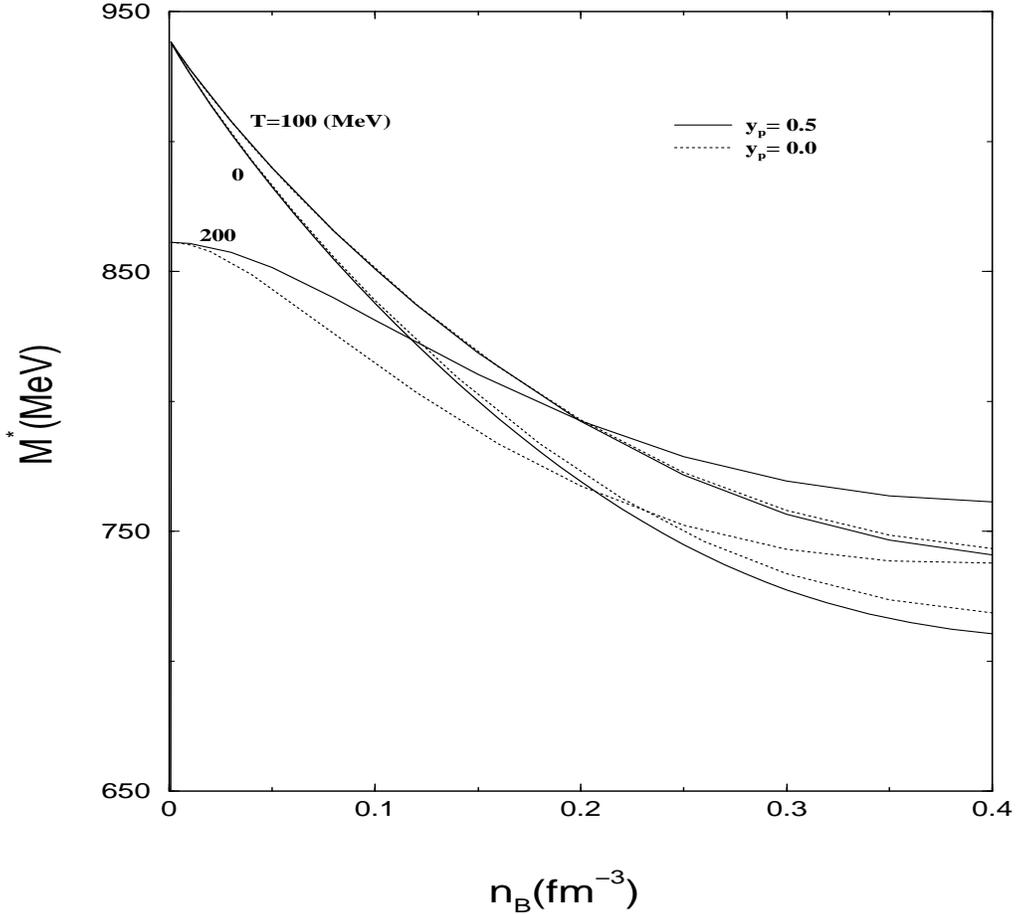}}
%\centerline{\epsfxsize=4in\epsfbox{pmu3.eps}}
\caption{\it{Effective mass as function of  baryon  number density
 at different temperatures.}} 
\end{figure}

    In Fig.2 we show the effective mass as a function temperature at  zero 
density($n_B$ = 0) and saturation density($n_B$ = 0.153fm$^{-3}$). The solid 
lines are for symmetric nuclear matter($y_p$=0.5) and doted lines for neutron 
matter($y_p=0$). It is clear from figure that for $n_B = 0$, the solid line 
and the dotted line are same, whereas for $n_B = n_0$, the $y_p$ dependence 
is sensitive to the temperature in between 150 MeV and 240 MeV and again the 
two lines coincide at higher temperatures. It is also observed that for 
$n_B = n_0$, the $M^*$ first increases slowly and then falls suddenly at 
about T $\approx$ 240 MeV. But for $n_B = 0$, $M^*$ remains almost constant as 
the temperature increases and falls suddenly at about T $\approx$ 235 MeV. This
result shows that a first order phase transition appears at $n_B = 0$, 
T $\approx$ 235 MeV which is similar to the result obtained for Walecka model, 
which has such a phase transition at T $\approx$ 185 MeV[16]. Because of 
strong attraction between the nucleons at  high temperatures the 
nucleon antinucleon pairs can be formed which may lead to abrupt change 
in $M^*$ to take place in high temperature region. But the mechanism of 
this first order phase transition is not clear. It is expected that this 
abrupt change of $M^*$ in the high temperature region for some models 
including the Walecka model might be related to the formation of new 
matter[17]. 
\begin{figure}[t]
\leavevmode
%\centerline{\psfxsize=2.5in\psfbox{pmu3.ps}}
\protect\centerline{\epsfxsize=5in\epsfysize=5in\epsfbox{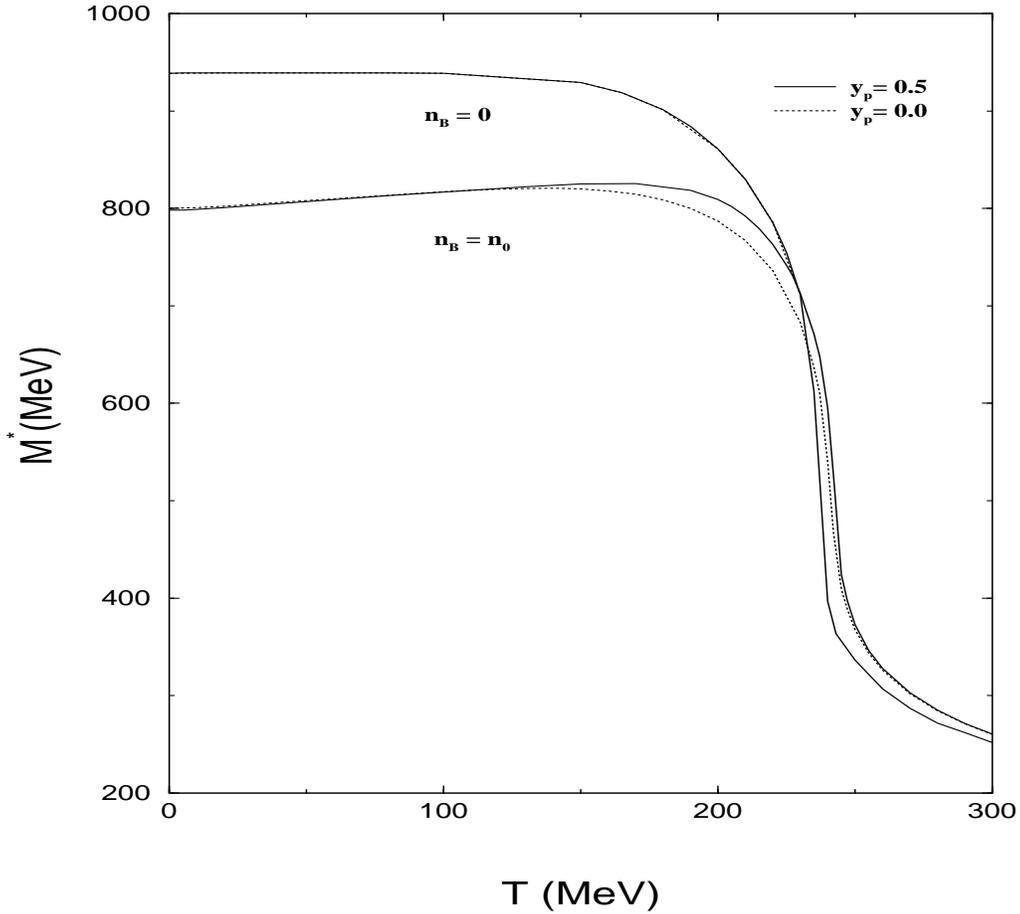}}
%\centerline{\epsfxsize=4in\epsfbox{pmu3.eps}}
\caption{\it{Effective mass as function of temperature for constant
  baryon number densities.}} 
\end{figure}

     Fig.3 shows the binding energy per nucleon as a function of the
baryon density at different temperatures for symmetric nuclear matter, 
$y_p$=0.5. At zero temperature it has a minimum at the nuclear saturation 
density $n_0$ which corresponds to a binding energy per nucleon of -16.3 MeV. 
With the increase of temperature the minimum shifts towards higher densities 
and for higher temperatures the minimum of the curve becomes positive. 
It is also observed that as the temperature increases,  
the nuclear matter  becomes less bound and the saturation curves in the 
MCH model are seems to be flatter than those observed in Walecka model[17,18]. 
This result implies that the nuclear matter EOS in MCH model is softer than 
that obtained in Walecka model.
\begin{figure}[t]
\leavevmode
%\centerline{\psfxsize=2.5in\psfbox{pmu3.ps}}
\protect\centerline{\epsfxsize=5in\epsfysize=5in\epsfbox{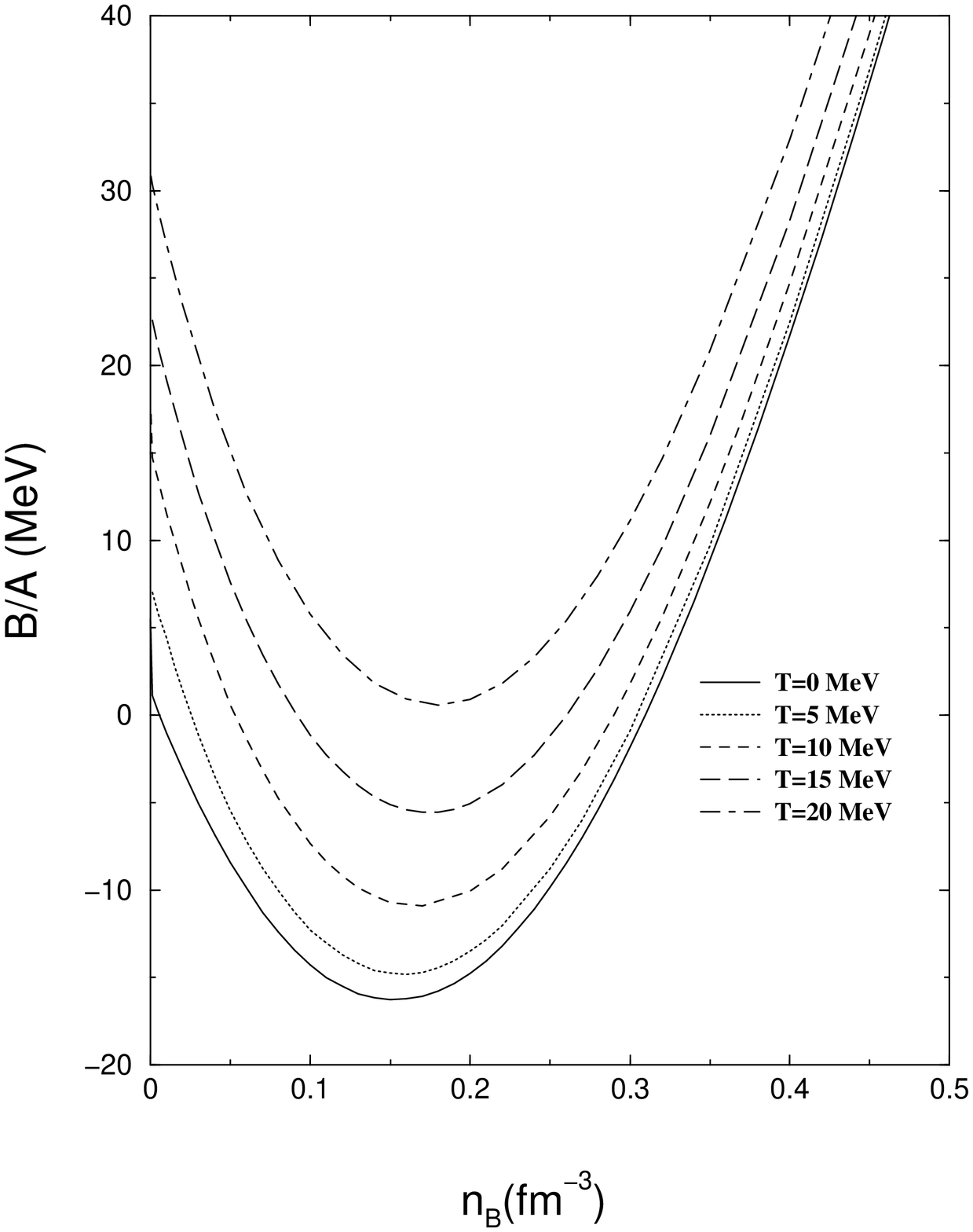}}
%\centerline{\epsfxsize=4in\epsfbox{pmu3.eps}}
\caption{\it{Binding energy per nucleon as function of  baryon  number density
 at different temperatures.}}
\end{figure}

   In Fig.4, we show pressure as a function of density $n_B$ at different 
 temperatures. The solid lines are for symmetric nuclear matter and the
dotted are for pure neutron matter which are stiffer at all temperatures.
For a given $n_B$, the pressure has the usual trend of 
increasing  with temperature[19]. As the temperature increases the EOS 
becomes stiffer. Pressure has a non-zero value for $n_B = 0$ at and
above  a temperature of 200 MeV. It indicates that pressure has a 
contribution arising from the thermal distribution functions for
baryons  and anti-baryons as well as from the non-zero
value of the scalar field. Similar results were also obtained by Panda 
$\it{et \ al.}$[20] for symmetric nuclear matter. The non-zero value 
for scalar $\sigma$ field has also been observed in Walecka model[19].
\begin{figure}[t]
\leavevmode
%\centerline{\psfxsize=2.5in\psfbox{pmu3.ps}}
\protect\centerline{\epsfxsize=5in\epsfysize=5in\epsfbox{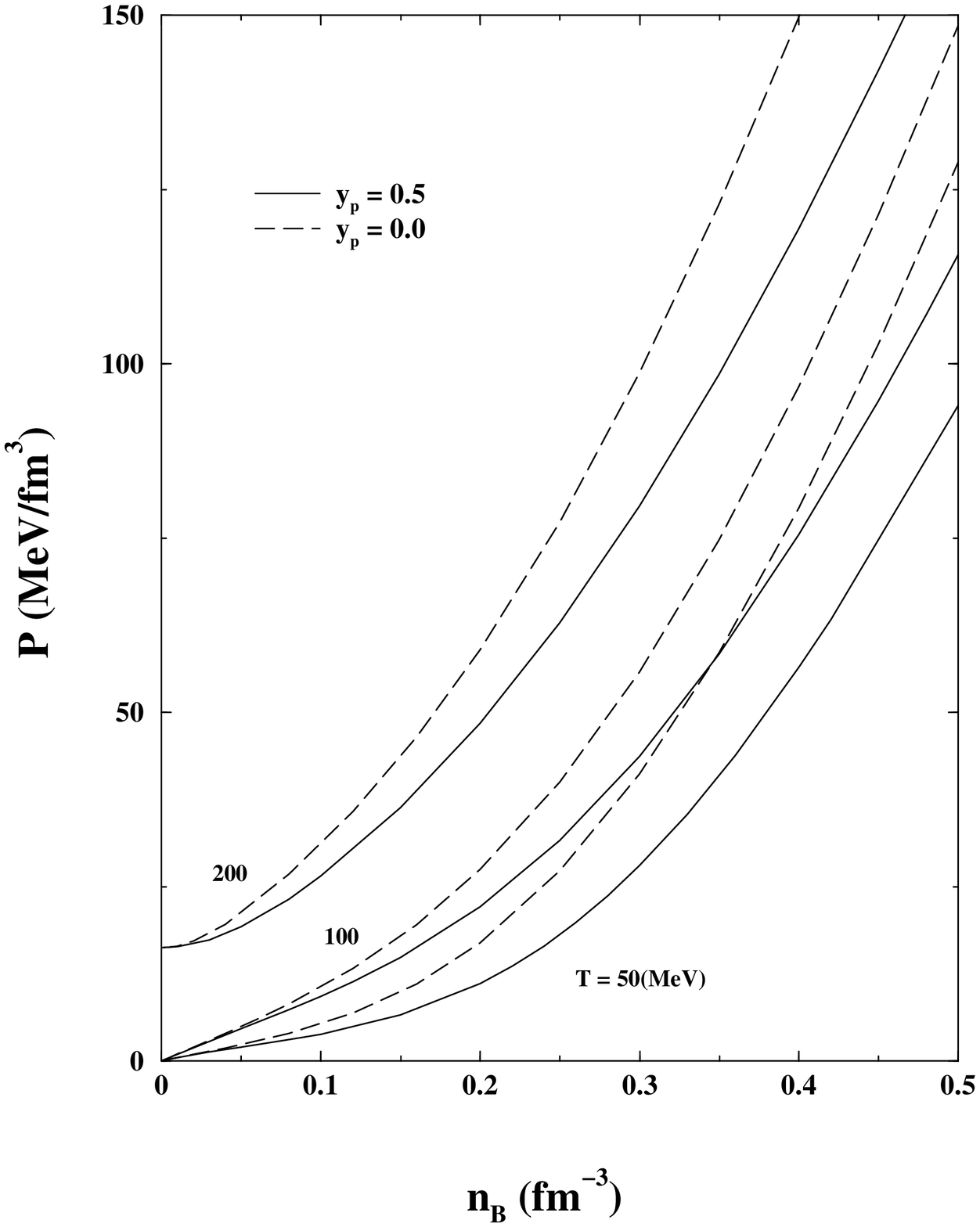}}
%\centerline{\epsfxsize=4in\epsfbox{pmu3.eps}}
\caption{\it{Pressure (P) as function of  baryon  number density
 at different temperatures.}} 
\end{figure}

     The entropy density as a function of density at different temperatures 
for symmetric nuclear matter and pure neutron matter is
presented in Fig.5. It is observed that entropy density for both is non-zero 
even at vanishing baryon density at a temperature of 200 MeV with 
contributions from the non-zero value of the sigma field. Similar behavior
was also observed for entropy density in the Walecka model and ZM model
calculations[18]. This increase of entropy density with increase of temperature
indicates a phase transition.
\begin{figure}[t]
\leavevmode
%\centerline{\psfxsize=2.5in\psfbox{pmu3.ps}}
\protect\centerline{\epsfxsize=5in\epsfysize=5in\epsfbox{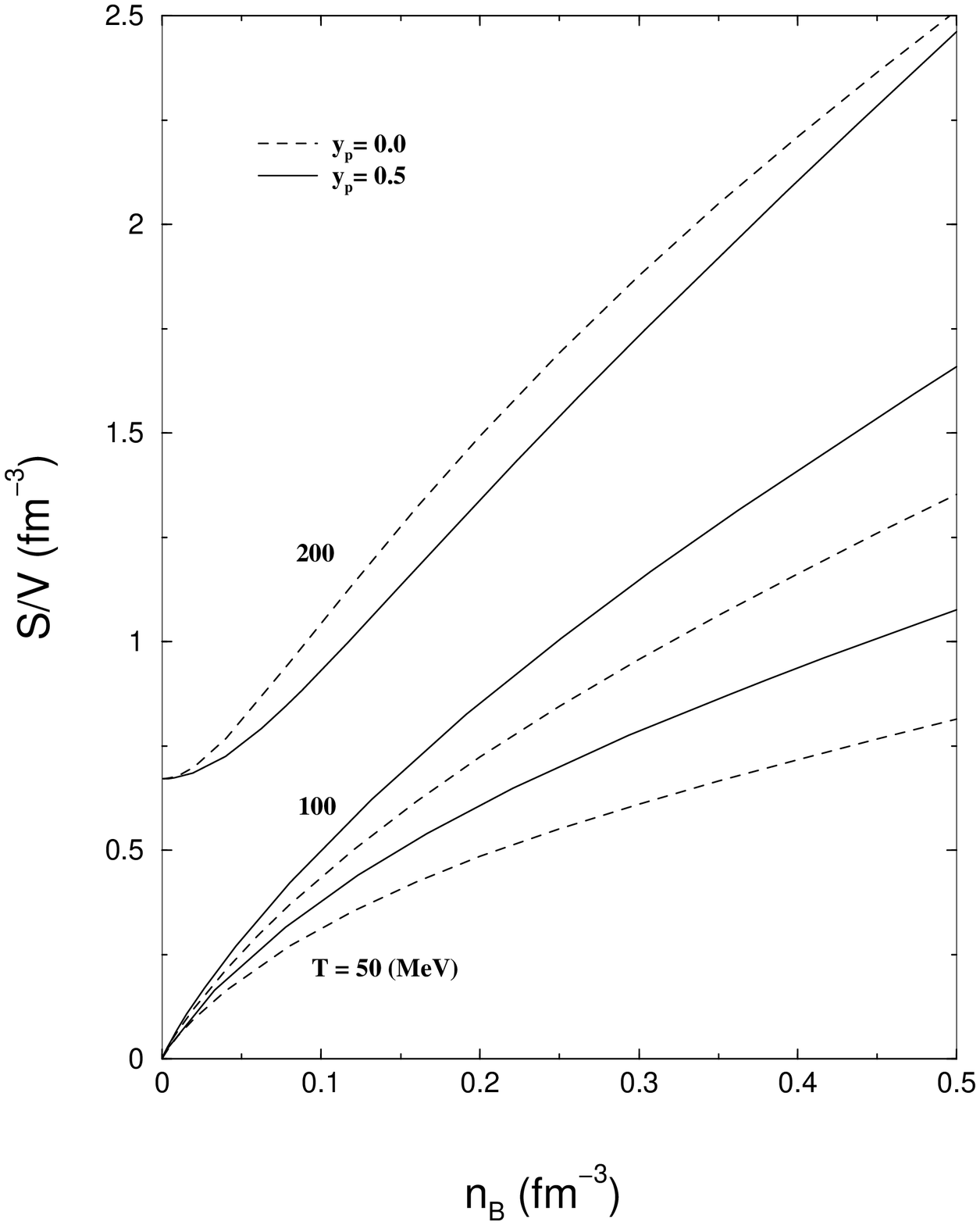}}
%\centerline{\epsfxsize=4in\epsfbox{pmu3.eps}}
\caption{\it{Entropy density as function of  baryon  number density
 at different temperatures.}}
\end{figure}
\noindent
The Entropy per baryon(S) as a function of  baryon  number density at 
different temperatures for symmetric nuclear matter is shown in Fig.6. At 
lower temperatures S decreases slowly as compared to higher temperatures and
the minimum value of S increases as the temperature increases which is 
similar to the result obtained in Refs.[2,8].
\begin{figure}[t]
\leavevmode
%\centerline{\psfxsize=2.5in\psfbox{pmu3.ps}}
\protect\centerline{\epsfxsize=5in\epsfysize=5in\epsfbox{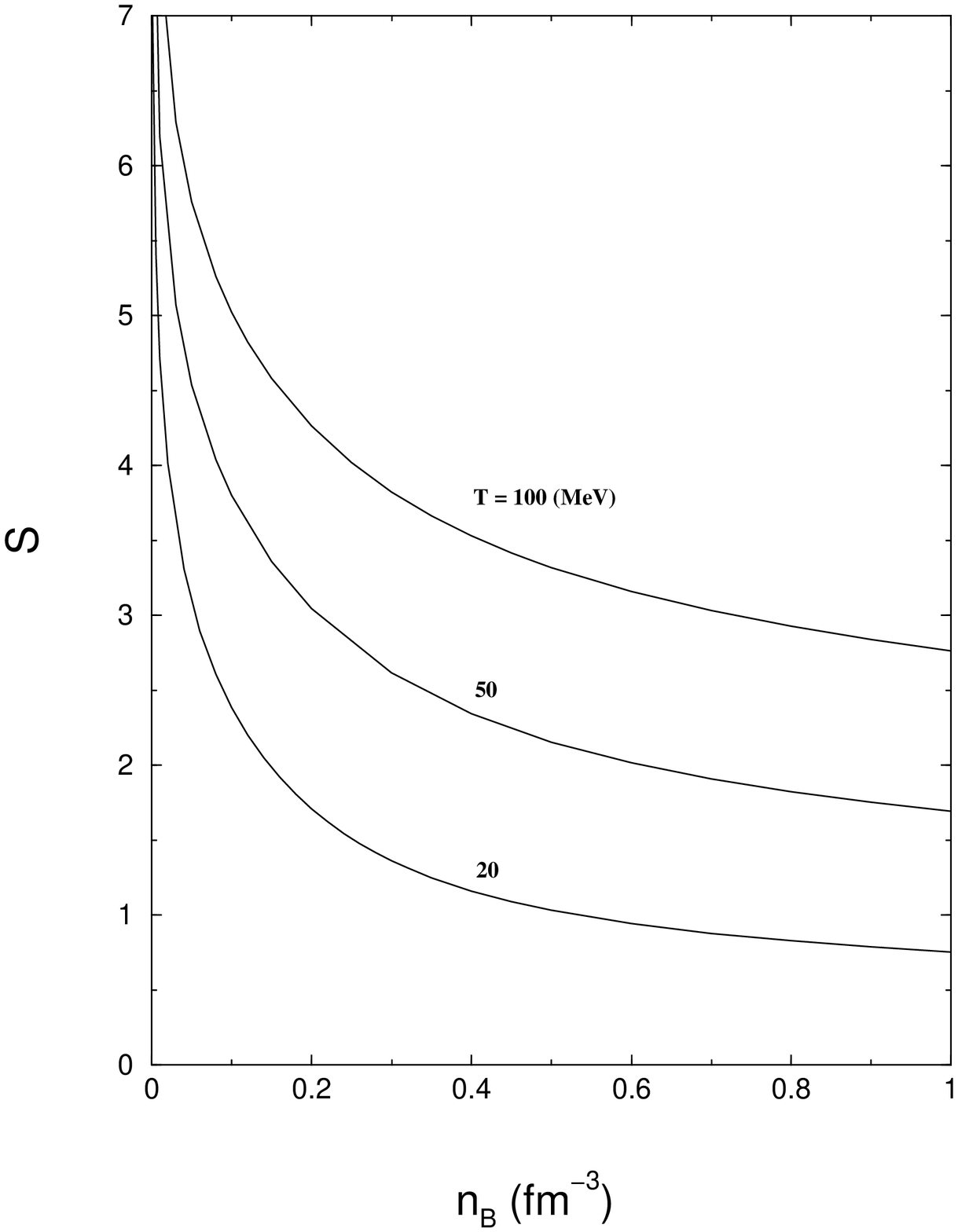}}
%\centerline{\epsfxsize=4in\epsfbox{pmu3.eps}}
\caption{\it{Entropy per baryon as a function of  baryon  number density
 at different temperatures.}}
\end{figure}

   We now  discuss the liquid-gas phase transition.
The pressure as a function of baryon density at fixed
temperature T = 10 MeV with different proton fractions is shown in Fig.7. 
It may be observed from the figure that for any fixed density with fixed 
T = 10 MeV, the pressure is not constant rather it increases with decrease of 
proton fraction. This clearly indicates that for asymmetric nuclear matter 
during isotherm liquid-gas phase transition the pressure can not remain 
constant but increases monotonically.  
It shows that for small $y_p$, particularly for neutron matter($y_p$=0), the
pressure increases monotonically which indicates that matter is stable 
at all densities. But for $y_p \geq 0.2$, the compressibility becomes 
negative, indicating mechanical instability. The diffusive unstable regions 
can be seen clearly from Fig.8, where the chemical potentials of proton and 
neutron is shown  as a function of $y_p$ at fixed pressure P = 0.1 MeV/fm$^3$ 
and temperature T = 10 MeV. According to the inequality[20] the region of 
negative slope for $\mu_p$ and positive slope for $\mu_n$ is unstable. Thus 
violation of stability criteria is an indication of phase separation.
\begin{figure}[t]
\leavevmode
%\centerline{\psfxsize=2.5in\psfbox{pmu3.ps}}
\protect\centerline{\epsfxsize=5in\epsfysize=5in\epsfbox{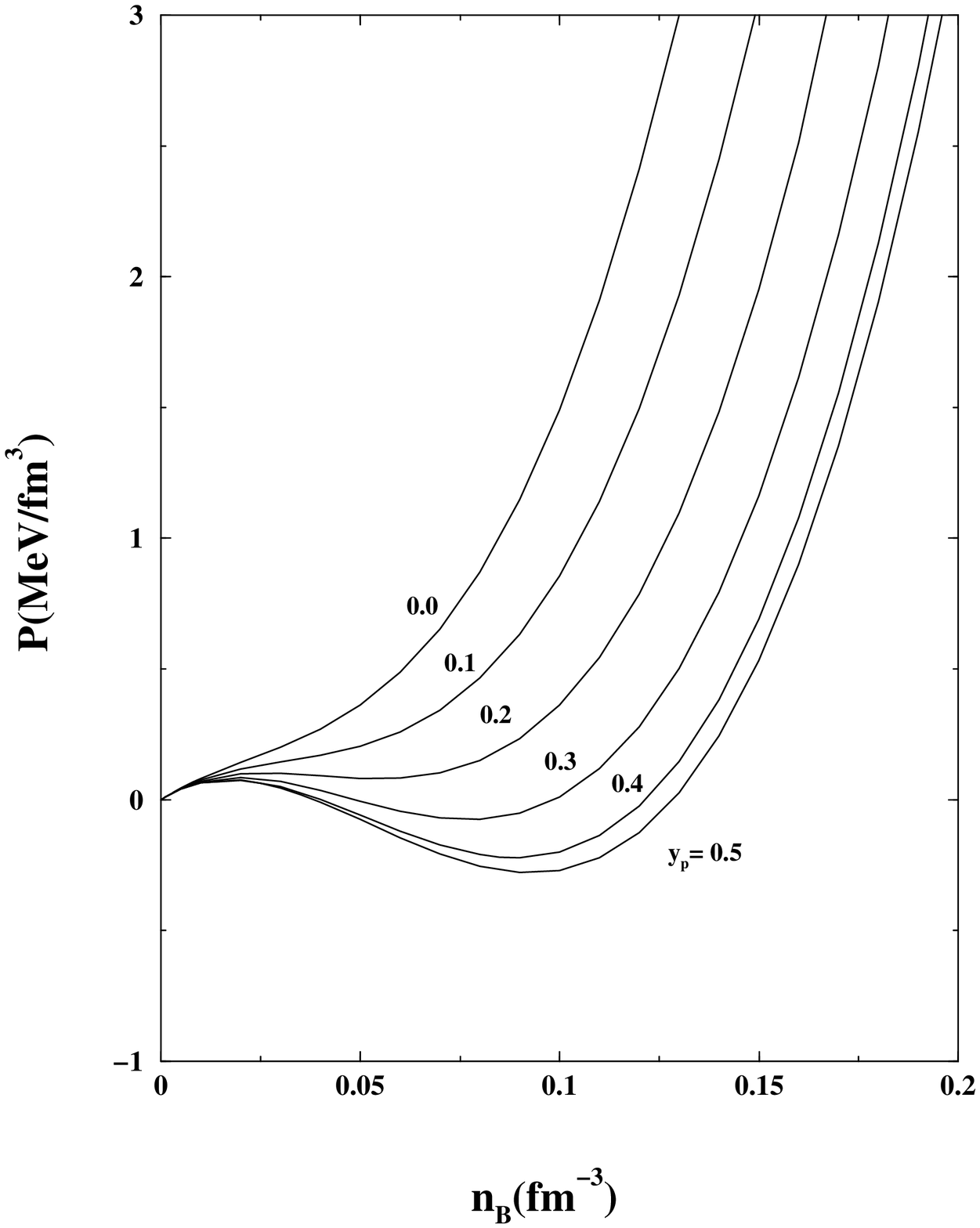}}
%\centerline{\epsfxsize=4in\epsfbox{pmu3.eps}}
\caption{\it{Pressure as a function of  baryon  number density for
    different proton fractions at temperature T=10 MeV.}}
\end{figure}

\begin{figure}[t]
\leavevmode
%\centerline{\psfxsize=2.5in\psfbox{pmu3.ps}}
\protect\centerline{\epsfxsize=5in\epsfysize=5in\epsfbox{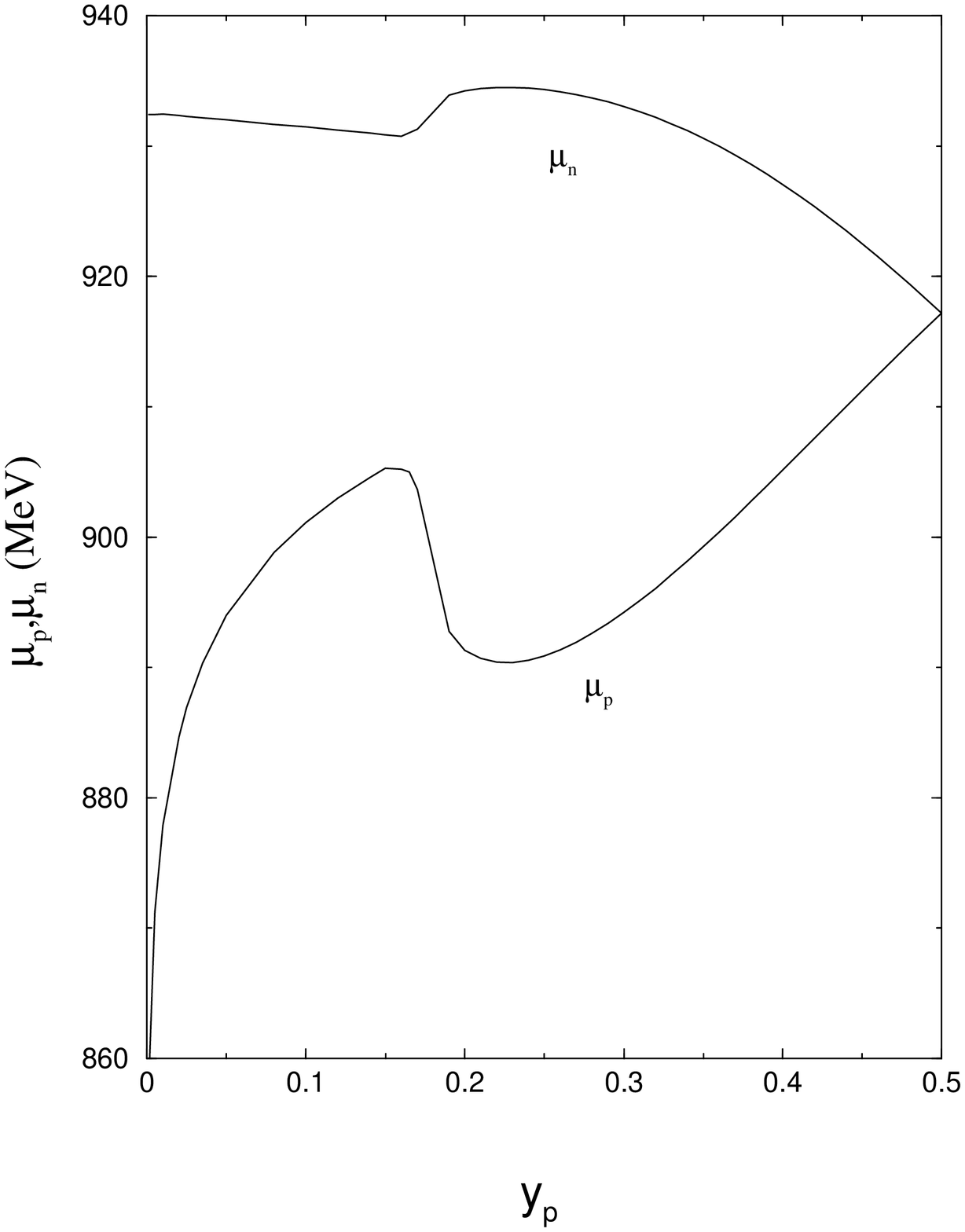}}
%\centerline{\epsfxsize=4in\epsfbox{pmu3.eps}}
\caption{\it{Chemical potentials as a function of $y_p$ at temperature 
    T=10 MeV and P=0.1 MeV/fm$^3$}}
\end{figure}

    Fig.9 shows the variation of pressure as a function of baryon
density for different $y_p$. One can see that the region of
mechanical instability decreases both with increase of temperature
and decrease of proton fraction[8]. 
\begin{figure}[t]
\leavevmode
%\centerline{\psfxsize=2.5in\psfbox{pmu3.ps}}
\protect\centerline{\epsfxsize=5in\epsfysize=5in\epsfbox{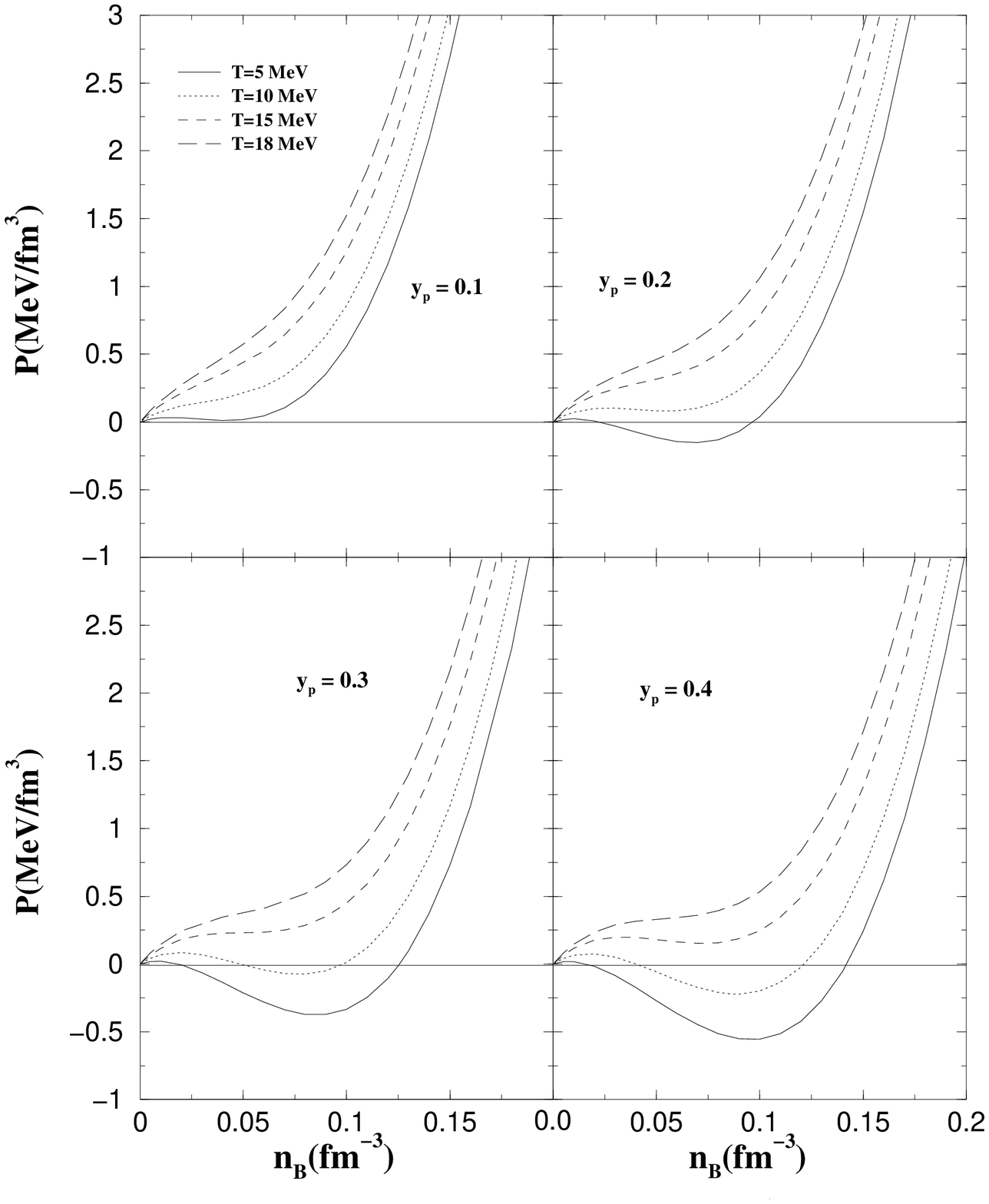}}
%\centerline{\epsfxsize=4in\epsfbox{pmu3.eps}}
\caption{\it{Pressure as a function of  baryon  number density for
    different proton fractions}}
\end{figure}
\noindent
We present the pressure of symmetric nuclear matter as a function of baryon
density for low temperatures in Fig.10. The figure shows that at zero
temperature, the pressure first decreases, then increases and passes through 
P = 0 at $n_B = n_0$(saturation density), where the binding energy per nucleon
is a minimum. Decrease of pressure with density implies a negative 
incompressibility, $K = 9(\frac{\partial P}{\partial n_B})$, which is a
sign of mechanical instability. When the temperature increases the region
of mechanical instability decreases and disappears at the critical 
temperature $T_c$, which is determined by $\frac{\partial P}
{\partial n_B}\mid_{T_c} = \frac{\partial^2 P}{\partial^2 n_B}\mid_{T_c} = 0$,
above which the liquid-gas phase transition is continuous. We have obtained 
the value of critical temperature $T_c \approx 17.2 MeV$, critical density   
$n_c \approx 0.045 fm^{-3} $,  critical pressure $ p_c \approx 0.274
MeV/fm^{-3}$ and critical effective mass $M^*_{c}\approx $ 887 MeV
which is in fair agreement with the results obtained in other studies[8,17,18].
\begin{figure}[t]
\leavevmode
%\centerline{\psfxsize=2.5in\psfbox{pmu3.ps}}
\protect\centerline{\epsfxsize=5in\epsfysize=5in\epsfbox{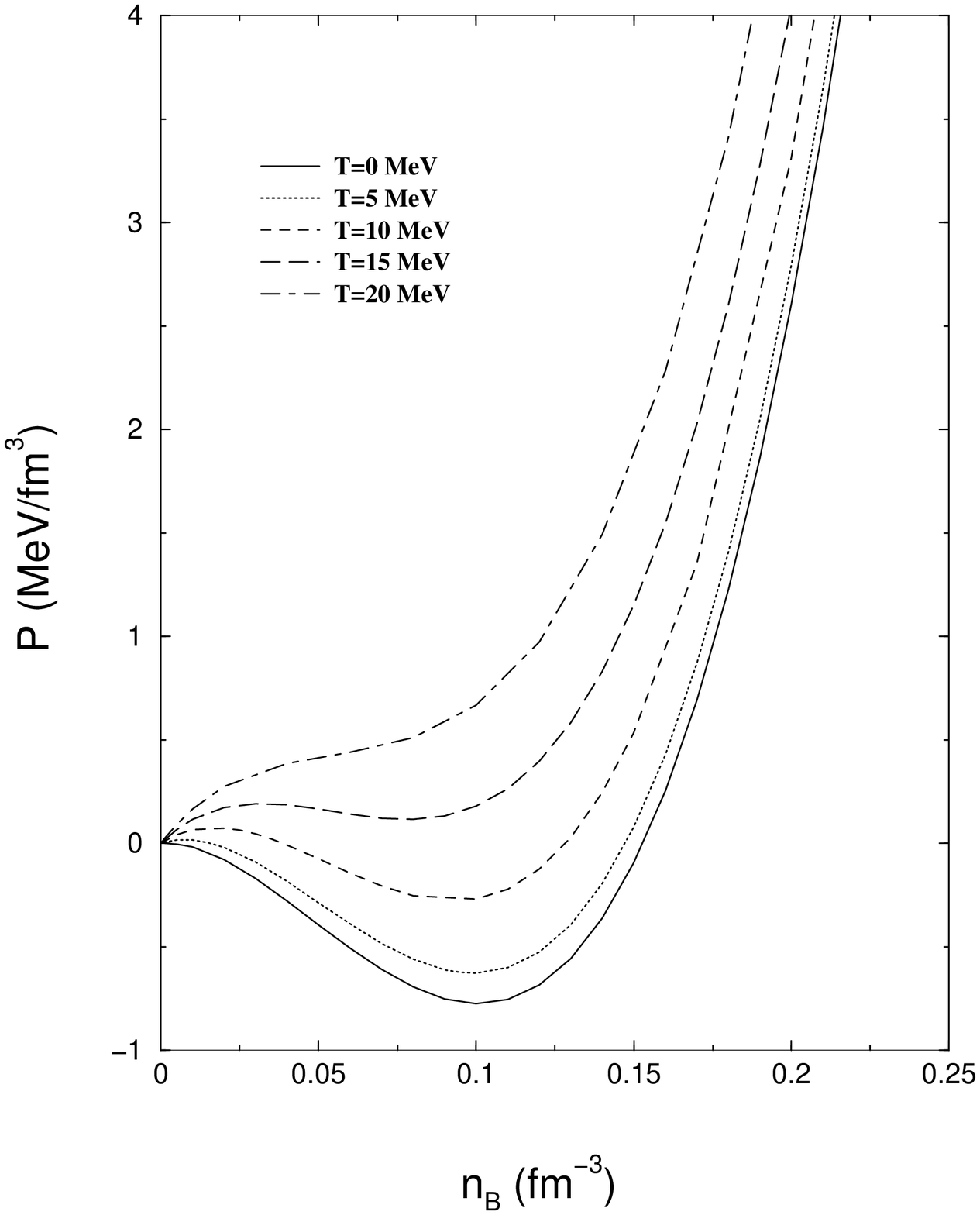}}
%\centerline{\epsfxsize=4in\epsfbox{pmu3.eps}}
\caption{\it{Pressure (P) as function of  baryon  number density
 for symmetric matter at different temperatures.}}
\end{figure}

   In Fig.11, we plot the variation of critical temperature with
different  proton fraction($y_p$) with and without $\rho$. The critical 
temperature $T_c$ decreases monotonically[8,21] as the proton fraction
decreases and goes to zero for $y_p=0.02$ with $\rho$ whereas
$T_c=11.6$ at $y_p=0$ without $\rho$. This indicates that addition of 
$\rho$-meson lowers the critical temperature at smaller $y_p$. As may be
seen from equations[8,9], the addition of $\rho$-meson gives a repulsive
potential which makes the nuclear matter easier to be gasified. But
for neutron matter($y_p=0$), the system only remain in gas phase even
at zero temperature[6]. 
\begin{figure}[t]
\leavevmode
%\centerline{\psfxsize=2.5in\psfbox{pmu3.ps}}
\protect\centerline{\epsfxsize=5in\epsfysize=5in\epsfbox{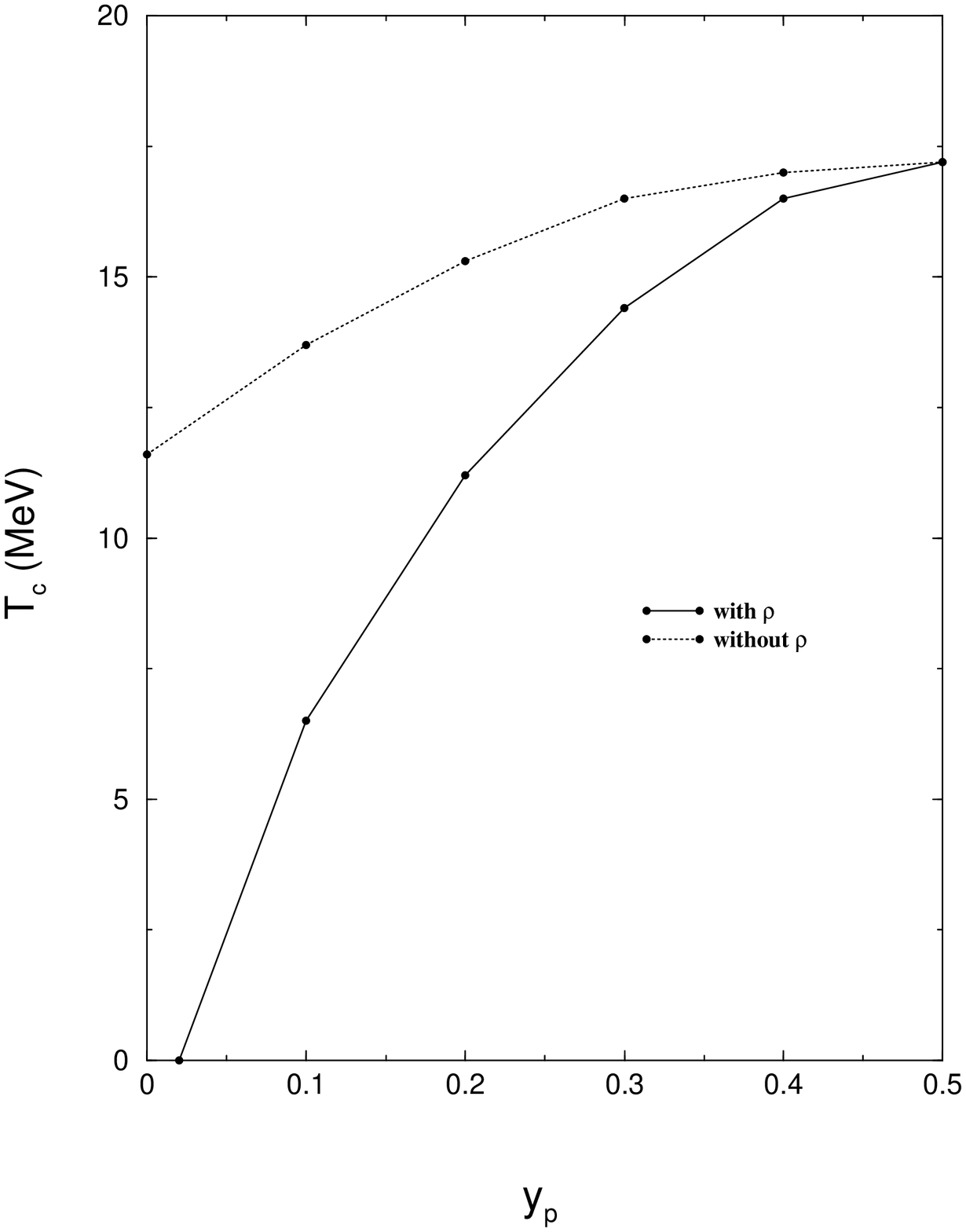}}
%\centerline{\epsfxsize=4in\epsfbox{pmu3.eps}}
\caption{\it{Critical temperature versus proton fraction $y_p$.}}
\end{figure}

\section{\bf{Proto-Neutron Star}}

  In this section we extend our investigation to a study of the structure and
properties of Proto-Neutron Star. A PNS is born
following the gravitational collapse of the core of a massive star
during a Supernova explosion(type-II). 
It is a hot collapsed core which can reach
temperatures as high as few tens of MeV. The evolution of PNS proceeding 
through several distinct states with various outcomes is discussed in
Ref.[2]. During the early evolution of PNS, a neutron star with an entropy per
baryon of order of unity contains neutrinos that are trapped in matter 
on dynamical time scale and after a lapse of few tens of seconds the
star achieves it's cold catalysed structure with essentially zero
temperature and zero trapped neutrinos. A PNS has approximately uniform
entropy per baryon(S) of 1-2 across the star[22]. At birth the PNS has
S=1. After deleptonization the entropy per baryon reaches its maximum
(S$\sim$ 2) and finally  cools down to its cold state with S=0[2].
The finite temperature aspect of EOS plays an important role in the
study of properties and structure of PNS. 
 
      The structure of PNS mainly depends on it's composition[2]. 
Since the composition of neutron star basically depends on the nature of
strong interactions which are not well understood in dense matter, one 
has to investigate various possible conditions taking many possible
models. Out of various possible cases discussed in Ref.[2], we consider
here  a case in which matter consists of neutrons, protons and
electrons whose relative concentrations is determined from the
conditions of charge neutrality and  $\beta$-equilibrium in the absence
of neutrino trapping[2].

    The $\beta$-equilibrium(without neutrino trapping) and the charge
neutrality conditions are respectively given by
\begin{equation}
\mu_n = \mu_p + \mu_e
\end{equation}
\noindent
and 
\begin{equation}
 n_p = n_e
\end{equation}
\noindent
   Where $\mu_e$ and $n_e$ are chemical potentials and number density
of electron respectively. The electron number density at finite
temperature can be written as
\begin {equation}
n_e = \frac{\gamma}{(2\pi)^3}\int_{0}^{\infty} d^3k \left[ n_{e}(T) 
 - \bar n_{e}(T)\right]
\end{equation}
\noindent
Where \\
\begin{equation}
  n_e(T) = \frac{1}{e^{(\sqrt(k^2 + m_e^2) - \mu_e )\beta} + 1} ,\ \ \ \ 
 \bar n_e(T) = \frac{1}{e^{(\sqrt(k^2 + m_e^2) + \mu_e )\beta} + 1} ,\\
\end{equation}
\noindent 
The number density of neutron and proton is defined in Eqns.12 and 13.
The extra terms which must be added to energy density and pressure
(given in eqns. 14 and 15) are respectively
\begin{equation}
\frac{\gamma}{2\pi^2}\int_{0}^{\infty} dk k^2\sqrt{(k^2+{m_e}^2)} 
   [n_e(T) + \bar n_e(T)] 
\end{equation}
\noindent
and 
\begin{equation} 
\frac{\gamma}{6\pi^2}\int_{0}^{\infty} \frac{dk k^4[n_e(T) + \bar n_e(T)]}
{\sqrt{(k^2+{m_e}^2)}}
\end{equation}
\noindent
For a given value of $n_B$, the equations[10,12,13,17,23] are to be
solved self-consistently using Eqns.19 and 20 (for fixed S=0,1 or 2) 
to get $\mu_p, \mu_e, n_p$, y, T  and then can be substituted 
to get pressure and energy density. After getting pressure as function
of energy density, the TOV equations can be integrated using proper
boundary conditions[13] to get mass and radius of star at fixed
entropy per baryon S. 

   Pressure as a function of number density for S = 0, 1 and 2 is shown in 
Fig.12. One can mark from the figure that the EOS becomes softer as the 
entropy per baryon decreases from S=2 to S=0 which indicates the lowering 
of mass and radius as shown in Fig.13 and Fig.14 respectively. For different 
values of S, the radius, energy density, pressure, number density and 
temperature corresponding to maximum mass is given in the following table.\\

\centerline{\bf{Table}}
\vspace{0.1in}
\centerline{
Star properties for matter in beta-equilibrium at finite entropy.}
\vspace{0.1in}

\begin{center}
\begin{tabular}{|c|c|c|c|c|c|c|}
\hline
S  & $\frac{M_{max}}{M_{\odot}}$ & R & $\epsilon_c$ & $n_c$ & $P_c$ & $T_c$ \\
 &  & (km) & ($\frac{MeV}{fm^3}$) &(fm$^{-3}$) & ($\frac{MeV}{fm^3}$)&(MeV) \\
\hline
0 & 2.18 & 12.14 & 1230 & 0.97 & 304.71 & 0.0 \\
1 & 2.21 & 12.23 & 1190 & 0.94 & 294.85 & 27.85 \\
2 & 2.33 & 12.45 & 1092 & 0.85 & 272.24 & 62.12 \\
\hline
\end{tabular}  
\end{center}

\vspace{0.4in}
The results in the table reflects the influence of entropy on the
gross properties of stars. Increase of maximum mass and radius upto
S=2, amounts to only a few percent of cold star and the temperature of 
the core is upto 62 MeV which is in fair agreement
with the results obtained in Ref.[2]. It also shows that for S=2, the
thermal contribution is larger which results in an increase of mass of 
about 0.15 $M_\odot$ compared with that for S=0.
In neutron stars, the pressure is supported by strongly interacting
baryons which have smaller contributions to the pressure which causes
a smaller increase in maximum mass. Thus the compositional variable of 
EOS has more importance than the temperature for the structure of
neutron star[2]. But for a White Dwarf which have highly ideal
EOS[23], the structure and properties are very  sensitivity to the
entropy per baryon.
\begin{figure}[t]
\leavevmode
\protect\centerline{\epsfxsize=5in\epsfysize=5in\epsfbox{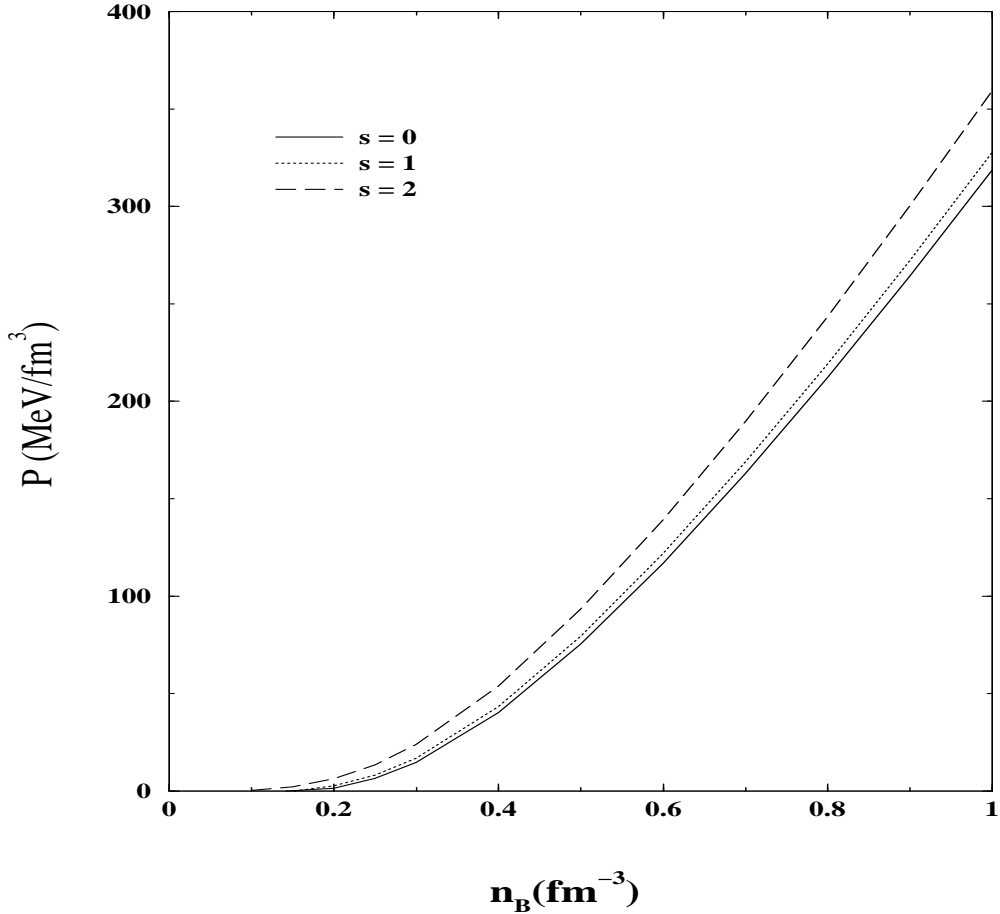}}
\caption{\it{Pressure as a function of number density at fixed entropy 
    per baryon.}}
\end{figure}

\begin{figure}[t]
\leavevmode
\protect\centerline{\epsfxsize=5in\epsfysize=5in\epsfbox{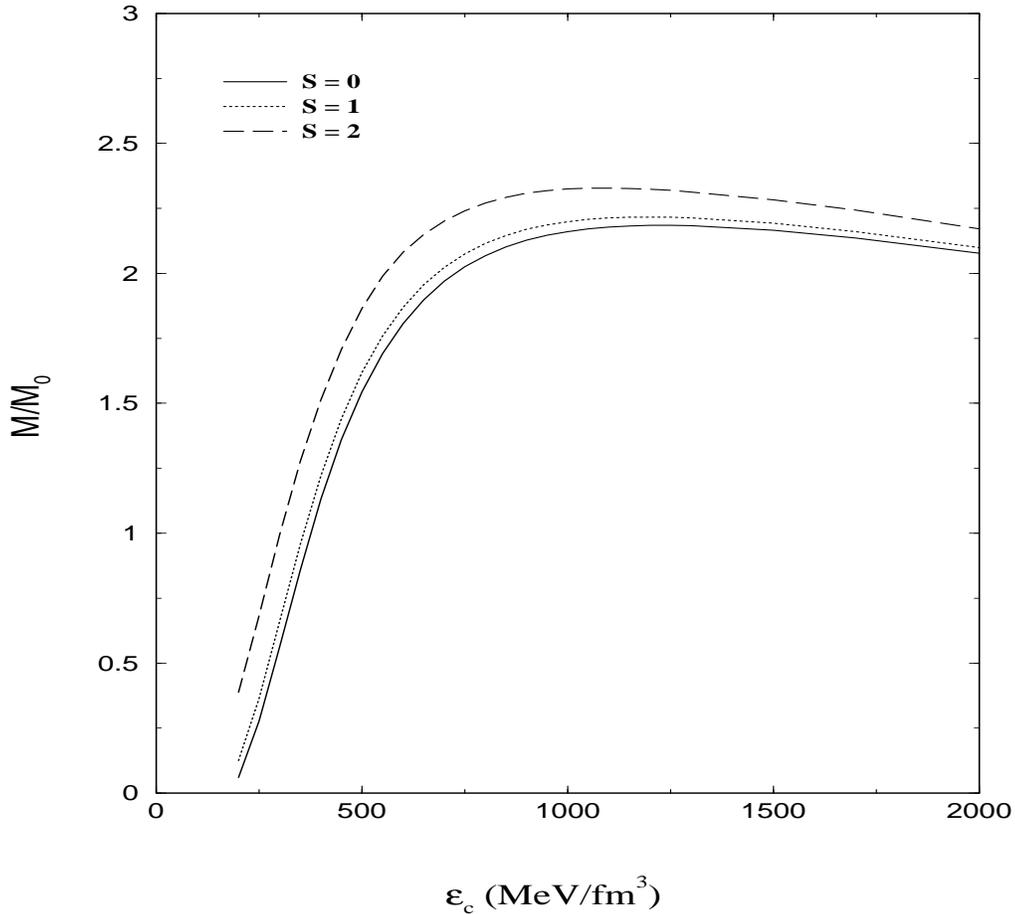}}
\caption{\it{Star mass (M/$M_{\odot}$) as a function of central energy
    density at fixed entropy per baryon.}}
\end{figure}

\begin{figure}[t]
\leavevmode
\protect\centerline{\epsfxsize=5in\epsfysize=5in\epsfbox{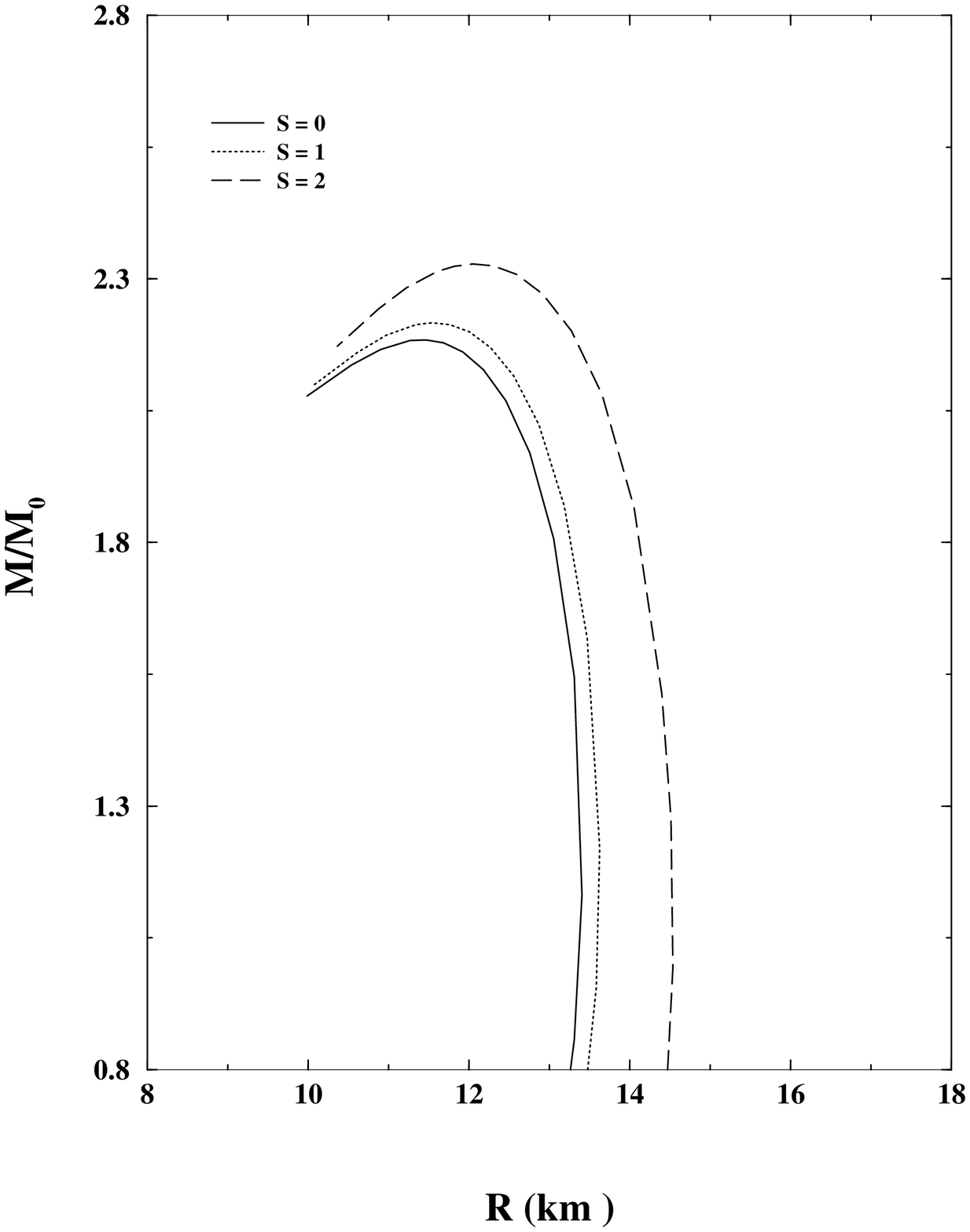}}
\caption{\it{Radius versus star mass at fixed entropy per baryon.}}
\end{figure}

    The relative concentrations of chemical potentials of n,p and e in
beta-equilibrium at a fixed entropy per baryon S=1, is shown in
Fig.15. It is clear that $\mu_e$ increases
linearly with number density whereas $\mu_n$ and $\mu_p$ first
decreases and then increases linearly. The increase of electron chemical 
potential with number density  implies the abundance of negatively 
charged particle(electron) which shows that the system is proton rich 
over an extended region of density. It may be seen from the figure
that in the very lower density region
the proton abundance is more, then decreases to some extent and then
increases linearly in high density region.
\begin{figure}[t]
\leavevmode
\protect\centerline{\epsfxsize=5in\epsfysize=5in\epsfbox{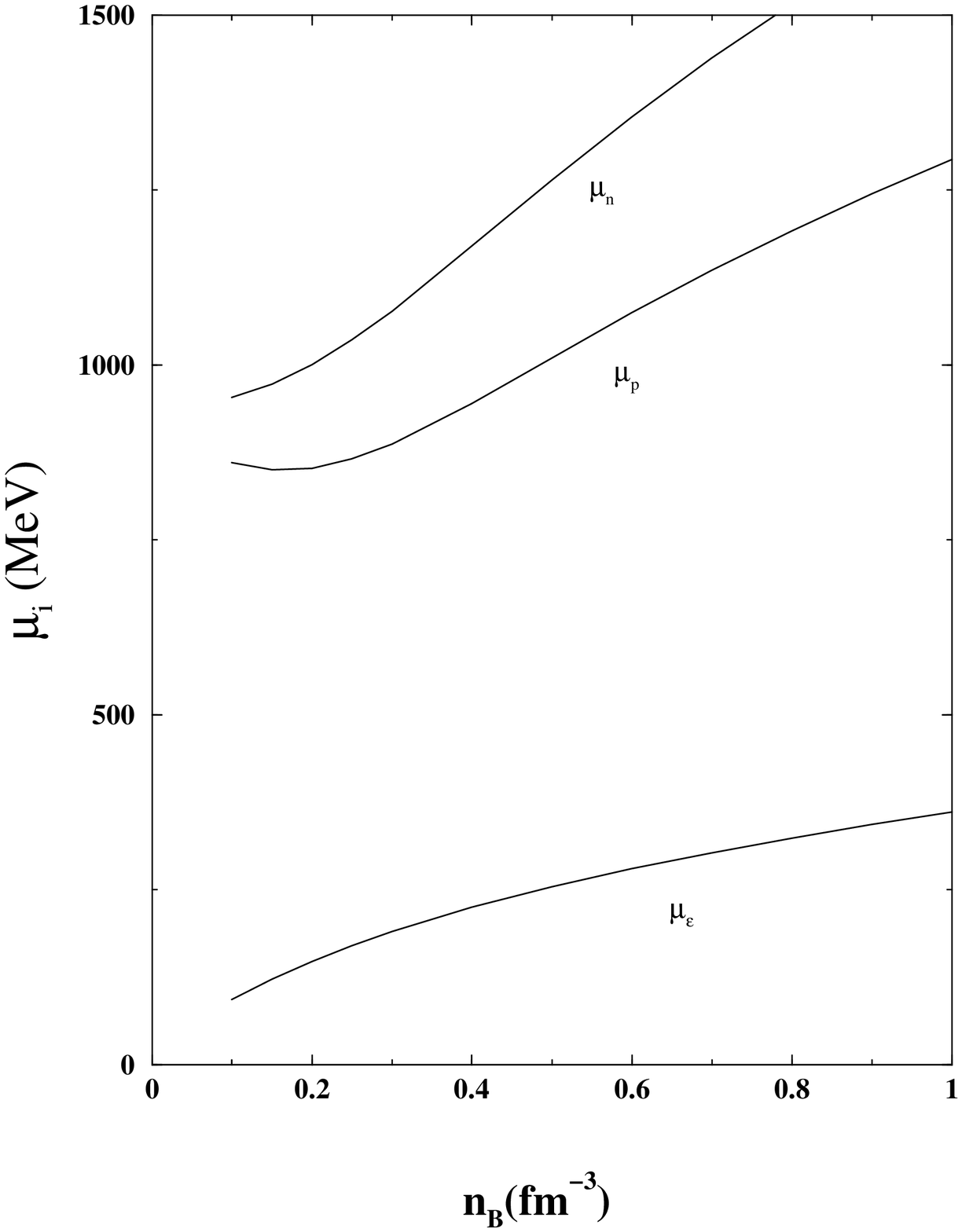}}
\caption{\it{Chemical potential versus number density in
    beta-equilibrium at fixed entropy per baryon S=1}}
\end{figure}

 The temperature as a function of energy density at fixed entropy per
baryon is shown in Fig.16. The temperature of the star increases for both
S=1 and S=2 from which one can get the critical temperature
corresponding to the maximum mass of the star.
The temperature is maximum at the center of 
the star (where central energy density is about 1100-1200 MeV/fm$^3$ for a
maximum mass star) and decreases with decreasing energy density which
is faster particularly at lower energy densities. This implies that
the interior of the star maintains a small variation of temperature
but  falls rapidly towards the surface region as density decreases.
\begin{figure}[t]
\leavevmode
\protect\centerline{\epsfxsize=5in\epsfysize=5in\epsfbox{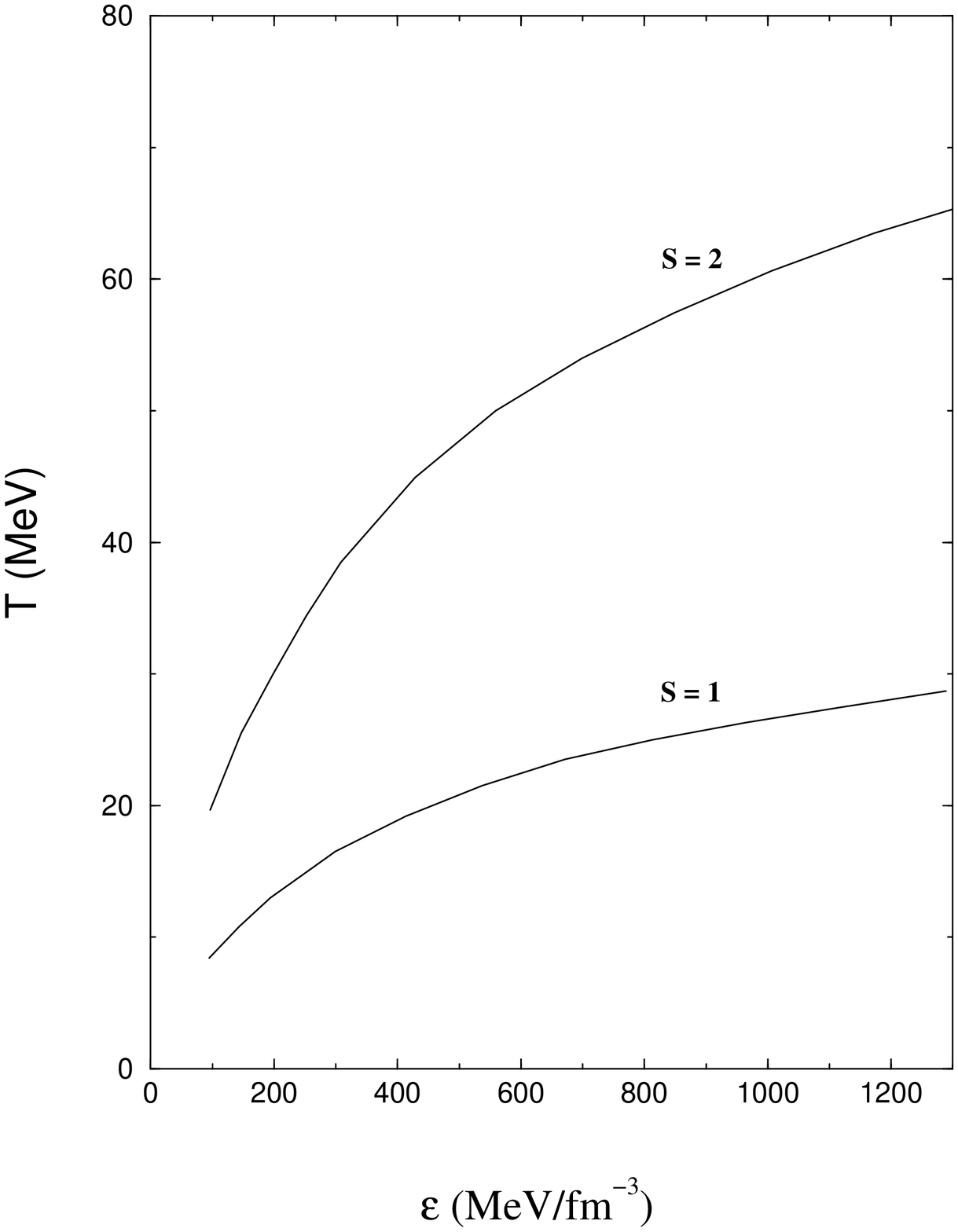}}
\caption{\it{Temperature versus energy density of proto-neutron star.}}
\end{figure}

\section{\bf{Conclusions}}

  We have studied here asymmetric nuclear matter at finite temperature using 
the MCH model[13]. We have presented how the effective nucleon mass, energy  
per baryon, entropy density, entropy per baryon and pressure behave as a 
function of density for various temperatures. At zero  density we 
find that this model exhibit a phase transition at T$\approx$ 235 MeV just as 
obtained in the Walecka model at T$\approx$ 185 MeV. 
This model exhibits the existence of liquid-gas phase transition in 
asymmetric nuclear matter and the  critical temperature $T_c$ depends 
sensitively on the proton fraction $y_p$. $T_c$ decreases with the decrease 
of $y_p$ as shown in Fig.11. The value of
critical temperature decreases from $T_c\approx $ 17.2 to
11.6 MeV for $y_p = 0.5$ to 0 without $\rho$-meson and from 
$T_c\approx $ 17.2 to 0 MeV for $y_p = 0.5$ to 0.02 with $\rho$-meson.
Hence the addition of $\rho$-meson seems to be very important to the 
study of properties of asymmetric nuclear matter as it lowers the critical 
temperature. This also shows that even at zero temperature the system  
remain only in gas phase for neutron matter($y_p = 0$). 
At fixed temperature and density the pressure of the system increases
with decrease of proton fraction(shown in Fig.7) which indicates that
during the isotherm liquid-gas phase transition pressure can not
remain constant for asymmetric nuclear matter.
We have also studied the EOS and structure of PNS with neutrino free charge
neutral matter in beta-equilibrium. We find that as the PNS cools from 
S = 2 to S = 0, the maximum mass and radius  exhibit a slow decrease. 
Thus the influence of entropy per baryon or equivalently
the temperature, on the structure of PNS is not very sensitive.  It is 
also observed that at finite entropy per baryon e.g. S=1, the star is 
proton rich over an extended region of density. The temperature varies 
slowly in the interior of the star but falls rapidly towards the low
density surface region and the maximum temperature of the core of the
star for S=2 is about 62 MeV. All these results of PNS are in fair
agreement with that obtained in Ref.[2].

%%**************************************************************************
\vspace {0.1in}
\noindent {\bf{Acknowledgements}}
\vspace {0.05in}

P.K.Jena would like to thank Council of Scientific and Industrial
Research, Government of India, for the award of SRF, with financial
support under the grand F.No. 9/173 (101)/2000/EMR-I. Help of the Institute of 
Physics, Bhubaneswar, India, is warmly acknowledged for providing the library 
and computational facility. 
%\newpage
%\section{\bf {References}}


\begin{thebibliography}{99}
 
\bibitem{ref1}E. Baron, J. Cooperstein and S. Kahana,
  Phy. Rev. Lett. $\bf{55}$,126 (1985); M. Brak, C. Guet and
  H. B. Hakansson, Phys.Rep.$\bf{123}$, 277 (1985).
\bibitem{ref2} M. Prakash, I. Bombaci, M. Prakash, P. J. Ellis, J. M. Lattimer 
  and R. Knorren, Phys. Rep. $\bf{280}$, 1 (1997).
\bibitem{ref3}W. A. Kupper, G. Wegmann and E. R. Hilf,
  Ann. phys.(N.Y.) $\bf{88}$, 454(1974); 
 B. Freedman  and V. R. Pandharipande, Nucl. phys. A  $\bf{361}$, 502(1981); 
 H. Jaqaman, A. Z. Mekjian and L. Zamick, Phy. Rev. C $\bf{27}$, 2782(1983);
 D. Bandyopadyay, C. Samanta, S. K. Samaddar and J. N. De, 
   Nucl. phys. A  $\bf{511}$, 1(1990).
\bibitem{ref4} H. Q. Song, Z. X. Qian and R. K. Su, Phy. Rev. C $\bf{47}$,
  2001(1993).
\bibitem{ref5} H. Q. Song  and R. K. Su, Phys. Lett. B $\bf{355}$, 179(1995).
\bibitem{ref6} P. Wang, Phy. Rev. C $\bf{61}$, 054904 (2003).
\bibitem{ref7} H. M\"{u}ller and B. D. Serot, Phy. Rev. C $\bf{52}$, 
  2072(1995).
\bibitem{ref8} P. K. Panda, G. Krein, D. P. Menezes and C. Providencia, 
   Phys. Rev. C $\bf{68}$, 015201(2003). 
\bibitem{ref9} B. D. Serot and J. D. Walecka, Adv. Nucl. Phys. $\bf{16}$ 1
  (1986); Int. J. Mod. Phys.E $\bf{6}$, 515 (1997).
\bibitem{ref10} J. A. Pons, S. Reddy, M. Prakash, J. M. Lattimer and
  J. A. Miralles, Astrophys.J. $\bf{513}$, 780(1999).
\bibitem{ref11} G.F.Marranghello, C.A.Z.Vasconcellos, M.Dillig and
  J.A.D.F.Pacheco, Int. J. Mod. Phys.E $\bf{11}$, 83(2002).
\bibitem{ref12} P. K. Jena and L. P. Singh, arXiv: nucl-th/0306085.
\bibitem{ref13} P. K. Jena and L. P. Singh, Mod. Phys. Lett. A   
   $\bf{17}$, 2633 (2002) ; Mod. Phys. Lett. A $\bf{18}$, 2135 (2003).
\bibitem{ref14} M. Barranco and J. R. Buchler, Phys. Rev.C  $\bf{22}$, 
  1729(1980).
\bibitem{ref15} N. K. Glendenning, Phys.Rev.D  $\bf{46}$, 1274 (1992).
\bibitem{ref16} J. Theis, G. Graebner, G. Buchwald, J. Maruhn,
  W. Greiner, H. St$\ddot{o}$cker, and J. Polonyi, Phys.Rev.D   
 $\bf{28}$, 2286 (1983).
\bibitem{ref17} Guo Hua, Liu Bo, M. Di Toro, Phys. Rev. C $\bf{62}$, 
  035203(2000).
\bibitem{ref18} M. Malheiro, A. Delfino and C. T. Coelho, Phys. Rev. C 
  $\bf{58}$, 426(1998). 
\bibitem{ref19} R. J. Furnstahl and B. D. Serot, Phys. Rev.C $\bf{41}$, 
  262(1990).
\bibitem{ref20} P. K. Panda, A. Mishra, J. M. Eisenberg and W. Greiner
  Phys. Rev. C  $\bf{56}$, 3134(1997).
\bibitem{ref21} S. Ray, J. Shamanna, T. T. S. Kuo, Phys. Lett. B $\bf{392}$,
  7(1997).
\bibitem{ref22} A. Burrows and J. M. Lattimer, Astrophys.J.$\bf{307}$, 
  178(1986).
\bibitem{ref23} J. I. Kapusta, $\it{ Finite \ Temperature \ Field \ Theory.}$
   (Cambridge Univ.Press).
\end{thebibliography}
\end{document}